\documentclass[prl,twocolumn,lengthcheck]{revtex4}

\begin{document}

\title{A simple operational interpretation of the fidelity}
\author{Jennifer L. Dodd}
\author{Michael A. Nielsen}
\affiliation{Centre for Quantum Computer Technology, Department of Physics,
University of Queensland 4072, Australia
\\Institute for Theoretical Physics, University of California, Santa Barbara,
CA 93106, USA}

\maketitle

{\bf This note presents a corollary to Uhlmann's
theorem~\cite{Uhlmann76,Jozsa94} which provides a simple operational
interpretation for the fidelity of mixed states.}

For any states $\rho$ and $\sigma$, the {\em fidelity} is defined by
$F(\rho,\sigma)\equiv{\rm tr}\sqrt{\rho^{1/2}\sigma\rho^{1/2}}$.
[See~\cite{Fuchs96} for other expressions for the fidelity and
extensive references.]  When applied to pure states $|\psi\rangle$ and
$|\phi\rangle$, this expression reduces to
$F(|\psi\rangle,|\phi\rangle)=|\langle\psi|\phi\rangle|$ which has an
operational interpretation in terms of the distinguishability of the
two states: when testing whether or not $|\phi\rangle$ is the same as
$|\psi\rangle$, the probability that $|\phi\rangle$ passes the test is
$F(|\psi\rangle,|\phi\rangle)^2$.  Unfortunately, such an
interpretation is lacking when the input states are mixed.  In this
note, we present a result which gives the mixed-state fidelity a
simple operational interpretation.  It is a corollary to Uhlmann's
theorem, which provides a formula for the fidelity in terms of
purifications:

\noindent{\bf Uhlmann's Theorem:} Let $\rho$ and $\sigma$ be two
states of a quantum system $Q$, and let $E$ be a second system with
dimension greater than or equal to the dimension of $Q$.  Then
\begin{equation} \label{Uhlmann}
F(\rho,\sigma)=\max|\langle\psi_0|\phi_0\rangle|
\end{equation}
where the maximization runs over all $|\psi_0\rangle$ and
$|\phi_0\rangle$ which are {\em purifications of $\rho$ and $\sigma$}
in $EQ$.

The corollary gives an alternative formula for the fidelity in which
the purifying systems of Uhlmann's theorem have been replaced by
quantum operations:

\noindent{\bf Corollary:} Let $\rho$ and $\sigma$ be two states of a
quantum system $Q$.  Then
\begin{equation} \label{new}
F(\rho,\sigma)=\max|\langle\psi|\phi\rangle|
\end{equation}
where the maximization runs over all pure states $|\psi\rangle$ and
$|\phi\rangle$ which are {\em taken to $\rho$ and $\sigma$ by some
quantum operation ${\mathcal E}$}, that is, ${\mathcal E}(\psi)=\rho$,
and ${\mathcal E}(\phi)=\sigma$.  [Note: a quantum operation is a
completely positive trace-preserving map~\cite{Nielsen00}, and we use
the shorthand ${\mathcal E}(\alpha)\equiv{\mathcal
E}(|\alpha\rangle\langle\alpha|)$.]

An interpretation suggested by this characterization is the following:
If $\rho$ and $\sigma$ are the (potentially mixed) output states of a
noisy channel, then the fidelity $F(\rho,\sigma)$ is a lower bound on
the overlap of the input states, assuming they were pure.

{\bf Proof of the Corollary:} Denote the quantity on the right-hand
side of Eq.~(\ref{new}) by $F'(\rho,\sigma)$.  We aim to show
$F'(\rho,\sigma)=F(\rho,\sigma)$ using Uhlmann's theorem.  The crucial
step is a relationship between quantum operations and purifications.
Suppose $Q$ is initially in the state $|\beta\rangle$, and ${\mathcal
E}$ takes $|\beta\rangle$ to $\tau$.  We can always introduce a second
system $E$ (assumed to be in the initial state $|0\rangle$, and with
dimension equal to the dimension of $Q$ squared) so that
\begin{equation}
\tau={\mathcal E}(\beta)={\rm
tr}_E\left[U|0\rangle|\beta\rangle\langle 0|\langle
\beta|U^{\dagger}\right]
\end{equation}
for some unitary $U$ which acts on $EQ$.  Comparing the two sides of
this equation, we see that $U|0\rangle|\beta\rangle$ purifies $\tau$
in the joint system $EQ$.

So, for every pair $|\psi\rangle$, $|\phi\rangle$, the condition
``$\exists{\mathcal E}:{\mathcal E}(\psi)=\rho$ and ${\mathcal
E}(\phi)=\sigma$'' can be rewritten as ``$\exists
U:U|0\rangle|\psi\rangle$ and $U|0\rangle|\phi\rangle$ purify $\rho$
and $\sigma$, respectively''.  Therefore, from Uhlmann's theorem, we
have
\begin{equation} \label{inequality}
F(\rho,\sigma)\geq\max|\langle 0|\langle\psi|U^{\dagger}U|0\rangle|\phi\rangle|=\max|\langle\psi|\phi\rangle|
\end{equation}
where the maximization runs over states $|\psi\rangle$, $|\phi\rangle$
satisfying the condition above.  But the quantity on the right-hand
side is just $F'(\rho,\sigma)$, so we have $F(\rho,\sigma)\geq
F'(\rho,\sigma)$.

Furthermore, every pair $|\psi_0\rangle$, $|\phi_0\rangle$ in the
composite space $EQ$ can be obtained by a unitary transformation on
a pair $|0\rangle_E|\psi\rangle_Q$, $|0\rangle_E|\phi\rangle_Q$,
provided $\langle\psi|\phi\rangle=\langle\psi_0|\phi_0\rangle$.
Therefore, the maximization in Eq.~(\ref{inequality}) runs over the
same values as the maximization in Eq.~(\ref{Uhlmann}), so we have
$F(\rho,\sigma)=F'(\rho,\sigma)$, which completes the proof.

The value of this corollary lies in providing an operational
interpretation for the fidelity.  Furthermore, this approach may be
valuable for suggesting new approaches to problems.  As an example,
this characterization suggests a simple, intuitive proof of the
well-known fact that no quantum operation can increase the
distinguishability of two quantum states~\cite{Barnum96}.  In terms of
the fidelity, this means that for any quantum operation ${\mathcal G}$
which acts on $Q$, $F({\mathcal G}(\rho),{\mathcal G}(\sigma))\geq
F(\rho,\sigma)$ for all states $\rho$ and $\sigma$ of $Q$.  The proof
is immediate from the corollary: Choose $|\psi\rangle$, $|\phi\rangle$
and $\mathcal{E}$ such that $\mathcal{E}(\psi)=\rho$,
$\mathcal{E}(\phi)=\sigma$ and
$F(\rho,\sigma)=|\langle\psi|\phi\rangle|$.  Then $\mathcal{G\circ E}$
takes $|\psi\rangle$ and $|\phi\rangle$ to $\mathcal{G}(\rho)$ and
$\mathcal{G}(\sigma)$, respectively.  Therefore, by the corollary,
$F({\mathcal G}(\rho),{\mathcal
G}(\sigma))\geq|\langle\psi|\phi\rangle|=F(\rho,\sigma)$.

{\bf Acknowledgment:} This research was supported in part by the
National Science Foundation under Grant No. PHY99-07949.

\end{document}